# Transition from reconstruction towards thin film on the (110) surface of strontium titanate

Z. Wang*,[1], A. Loon[2], A. Subramanian[2], S. Gerhold[1], E. McDermott[3], J. A. Enterkin[4], M. Hieckel[1], B. C. Russell[5], R. J. Green[6], A. Moewes[6], J. Guo[7], P. Blaha[3], M. R. Castell[5], U. Diebold[1], and L. D. Marks[2]

[1]Institute of Applied Physics, TU Wien, Wiedner Hauptstrasse 8-10/134, 1040 Vienna, Austria
[2]Department of Materials Science and Engineering,
Northwestern University, Evanston, IL 60208, USA
[3]Institute of Materials Chemistry, TU Wien, Getreidemarkt 9/165-TC, 1060 Vienna, Austria
[4]Department of Chemistry,
Northwestern University, Evanston, IL 60208, USA
[5]Department of Materials, University of Oxford,
Parks Road, Oxford OX1 3PH, UK
[6]Department of Physics and Engineering Physics, University of Saskatchewan,
116 Science Place, Saskatoon, SK S7N 5E2, Canada
[7]Beijing National Laboratory for Condensed Matter Physics and Institute of Physics, Chinese Academy of Sciences, Beijing 100190, People's Republic of China

**Abstract**

The surfaces of metal oxides often are reconstructed with a geometry and composition that is considerably different from a simple termination of the bulk. Such structures can also be viewed as ultrathin films, epitaxed on a substrate. Here, the reconstructions of the $SrTiO_3$ (110) surface are studied combining scanning tunneling microscopy, transmission electron diffraction, and X-ray absorption spectroscopy, and analyzed with density functional theory calculations. While $SrTiO_3$ (110) invariably terminates with an overlayer of titania, with increasing density its structure switches from n×1 and 2×n. At the same time the coordination of the Ti atoms changes from a network of corner-sharing tetrahedra to a double layer of edge-shared octahedra with bridging units of octahedrally coordinated strontium. This transition from the n×1 to 2×n reconstructions is a transition from a pseudomorphically stabilized tetrahedral network towards an octahedral titania thin film with stress-relief from octahedral strontia units at the surface.

**Keywords:** Surface Structure, Reconstruction, Thin Film Nanostructures, Epitaxy, DFT

The increasing interest in SrTiO$_3$ as a functional ceramic material is due to its broad variety of optical, electrical, and chemical properties. The surfaces and interfaces are particularly important in the field of oxide electronics where it is used both as a substrate and an active material (e.g.[1-4]), and the electronic properties can be modulated through tailored epitaxial film growth (e.g.[5-7]). The surfaces also have interesting properties for applications such as model catalytic supports.[8-10] Although strontium titanate is the prototype perovskite, its surface structures make most other materials seem simple with a very large number of different reconstructions on the (001)[11] and (111)[12] surfaces, many of which are unsolved.

The surface reconstructions of simple binary oxides under oxidizing conditions involve only atomic rearrangements, but ternary oxides such as SrTiO$_3$ are more complex. There is an additional degree of freedom associated with changes in the surface composition, and all the reconstructions solved to date are known to be TiO$_2$ rich. The (110) surface of SrTiO$_3$ displays a range of different reconstructions (see[13] and references therein). The first set of structures to be decoded was a family of n×1 surface reconstructions containing corner-shared tetrahedral TiO$_4$ units arranged in a homologous series of valence neutral reconstructions.[13] The structural details have been refined by STM.[14] While many features are now known, for instance the role of strontium at domain-boundaries,[15] these are not the only structures that form on this surface. Of particular relevance here, the n×1 reconstructions can coexist with small regions of 1×n surface structures, which STM images indicate occur near surface steps.

The n×1 reconstructions are not simply a minor adjustment of the atomic positions, instead there is a complete change in the local titanium co-ordination. The local band-gap is substantially larger, all local properties will be quite different so they should be considered as a monolayer of a different phase at the surface, i.e. heteroepitaxial growth of an oxide on an oxide. A subtle issue is how does the surface/thin film structure develop as the surface excess of TiO$_2$ increases towards the limit at high excess of a heteroepitaxial TiO$_2$ thin film. Does the initial one or two atomic layer reconstruction template further layers; does the (110) surface transform back from the unusual tetrahedral co-ordination of the n×1 structures to a more standard octahedral structure? It is established that local bonding considerations matter for the monolayer thickness surface reconstructions,[13, 16] but as the excess TiO$_2$ increases will the system start to be dominated by the energetics of conventional epitaxial growth theory with, for instance, misfit dislocations and stress relief rather than local bonding dominating the thermodynamics?

These issues have recently become of increasing interest for oxides following the observation that for optimal growth of certain oxides by molecular-beam epitaxy (MBE) the deposition order that is required is different to that of the final desired material.[17-19] This is analogous to many other types of growth, for instance conventional MBE of III-V semiconductors where the best results require a surface rich in the V compound and the excess acts as a surfactant. A similar role for surfactants has also been recently found for the SrTiO$_3$ (001) surface where optimal growth of low defect material by hybrid MBE occurs when in reflection high-energy electron diffraction a c(4×4) structure is present during growth,[20] this reconstruction being one which contains chemisorbed water[21, 22] which presumably acts as the surfactant. While phenomena such as surfactant controlling growth are well established for epitaxial growth, one does not often consider a surface reconstruction as a surfactant – but clearly the c(4×4) acts as one and this could be a general phenomena. For oxides, are surface structures, epitaxy and growth always intimately linked or is there an evolution from one to another in controlling the final material?

We report here a step towards a better understanding of how oxide systems evolve from the monolayer to multilayer, from monolayer surface reconstruction towards multilayer heteroepitaxial thin film. This is done through an analysis of the high TiO$_2$ coverage 2×n reconstructions on the (110) SrTiO$_3$ surface from scanning tunneling microscopy, transmission electron diffraction and density functional theory calculations complemented by x-ray absorption spectroscopy. The adjustment of surface structures is performed in two ways: by vapor-depositing appropriate amounts of Ti and Sr, and annealing in O$_2$ at moderate temperatures as reported earlier [23], and by simple oxygen anneal at high temperatures. While the STM images of the external surface imply a structure similar to lepidocrocite titania nanosheets,[24-26] electron diffraction data show that it is more complicated with partial occupancy and disorder of the second-layer titanium sites, which is a conclusion verified by DFT energetics. The surface switches from tetrahedral TiO$_4$ coordination to a double layer of edge-shared TiO$_6$ octahedra with bridging units of octahedral strontium, which play the role of stress relaxation defects, with the transformation dominated by stress considerations. This transition towards more conventional epitaxial growth occurs for about 2.75 monolayers excess of TiO$_2$. There is still strain in the system with stress relief taking place at the top surface rather than at the interface, with issues of local bonding also important. The data indicate that it will only be after more than 3.5 monolayers that the system will start to approach a conventional thin film structure with conventional epitaxial terms dominant.

**Methods**

The $SrTiO_3$ (110) samples were studied in several different labs by various techniques. The two types of surface reconstructions, n×1 and 2×n, could be reproducibly obtained using two different approaches; these are referred to as low-temperature and high-temperature samples.

The high-temperature samples were used for the TEM studies as well as some STM imaging. As described previously[13], they were prepared by ion-beam sputtering followed by annealing in the temperature range 950-1100 °C in one atmosphere of $O_2$ for ~ 1 hr as well as as part of a more extensive study as a function of time and temperature.[27] In all cases high purity single crystal $SrTiO_3$ (110) wafers (MTI Corporation) were prepared using mechanical polishing and dimpling then annealed with slower cooling rates (~0.25 K/sec). Electron diffraction data were collected using a Hitachi H8100 at 200 kV, with the intensities extracted using the edm code.[28]

The low-temperature samples were used for the STM (Fig.1) and for the XAS measurements, and were checked with low-energy electron diffraction (LEED) in both cases (see Supplemental Figure S1). Nb-doped (0.5 wt%) $SrTiO_3$ (110) single crystal samples were purchased from MaTecK, Germany. In the low-temperature approach, the surfaces were initially sputtered and annealed at moderate temperature. By further deposition of Sr or Ti, followed by annealing at moderate temperatures in $O_2$, the surface structure could be precisely tuned according to the surface phase map shown in Figure S2 in Supplementary Information and ref.[23]. Here the surface stoichiometry, and thus structure, was adjusted until a desired mono-phased surface was observed via STM, and/or a sharp low-energy electron diffraction (LEED) pattern was present.[29,30] For the results shown in Fig. 1 the clean 4×1 surface was obtained by cycles of $Ar^+$ sputtering followed by annealing in $2\times10^{-6}$ mbar oxygen at 900 °C for 1 hr,[31] and the 2×5 surface was obtained h by depositing Ti at 600-700 °C in $6\times10^{-6}$ mbar $O_2$. STM images were obtained in a SPECS UHV system with a base pressure of $1\times10^{-10}$ mbar. XAS was performed at room temperature at the I311 beamline at Max-lab [32], structures were ascertained by LEED. A Scienta electron analyzer was used to measure the integrated intensity of the surface-sensitive Ti $L_{M,M}$

Auger lines with the detector in a surface-grazing orientation, as well as the more bulk-sensitive total secondary electron yield. The Ti $L_{2,3}$ spectra were analyzed using the CTM4XAS code.[33]

Remarkably, both the high- and the low-temperature approach give the same types of surface reconstructions. This is clear from the evolution of the electron diffraction patterns, and also from the fact that the STM images of 4×1 structure, observed on low-temperature samples, are in remarkably good agreement with the structural model proposed based on TEM results obtained for high-temperature samples. The XAS data shown below further confirm the 4×1 model, which is based on tetrahedral building blocks as a structural motif. In addition, STM data described in [34], which was collected using higher temperature annealing for comparable times, are comparable to our low-temperature STM results.

DFT calculations were performed with the all-electron augmented plane wave + local orbitals WIEN2K code.[35] The surface in-plane lattice parameters were those of the DFT optimized bulk cell, with ~1.2 nm of vacuum. STM images were simulated using the Tersoff-Hamann approximation.[36] Technical parameters were muffin-tin radii of 1.6, 2.36 and 1.75 Bohrs for O, Sr and Ti respectively, a min(RMT)*$K_{max}$ of 6.25, Brillouin-zone sampling equivalent to a 6×4 in-plane mesh for a (110) 1×1 cell with a Mermin functional. The electron density and all atomic positions were simultaneously converged using a quasi-Newton algorithm[37] to better than 0.01 eV/1×1 surface cell. Crystallographic Information Format (CIF) files for all the converged structures are in the Supplemental Material. The PBEsol[38] generalized gradient approximation as well as the revTPSSh method[39] were used, with 0.5 on-site exact-exchange based on earlier work.[40] The surface enthalpy per (1×1) surface unit cell ($E_{surf}$) was calculated as: $E_{surf}$=($E_{slab}$-$E_{STO}$*$N_{STO}$− $E_{TO}$*$N_{TO}$)/(2*$N_{1×1}$), where $E_{slab}$ is the total enthalpy of the slab, $E_{STO}$ for one bulk SrTiO$_3$ unit cell, $N_{STO}$ the number of bulk SrTiO$_3$ unit cells, $E_{TO}$ bulk rutile TiO$_2$, $N_{TO}$ the number of excess TiO$_2$ units and ($N_{1×1}$) the number of (1×1) cells. Consistency checks indicated an error in the energies of approximately 0.1eV/1×1 cell (~76 mJ/m$^2$, 10 kJ/mole).

**Results**

**Surface structural transition.** Depending upon the surface cation coverage, the SrTiO$_3$ (110) surface exhibits two distinct series of reconstructions, n×1 and 2×n. These reconstructions were obtained reversibly by depositing Ti/Sr followed by annealing (see Figure 1 and Supplemental

Figure S2), and the two reconstructions can coexist.[13] The n×1 structures are a homologous series composed a single layer of tetrahedrally coordinated TiO$_4$ residing directly on bulk truncated SrTiO$_3$ (110),[13] appearing as quasi-one-dimensional stripes along the [1$\bar{1}$0] direction (see the STM image of the 4×1 in Figure 1a). The 2×n (n = 4,5) structures show wider stripes along the [001] direction, separated by trenches with bright protrusions often, but not always, in every other row, leading to a 2× periodicity along the [001] direction. We note for later that this doubling was not observed for the STM images of samples prepared at high temperatures. Excluding the trench structure, the STM images are similar to published data for lepidocrocite titania nanosheets.[24-26]

The possibility exists that for some tip conditions the (0,0,2) repeat might not be observable, and appears as a smeared (0,0,1) repeat. A careful analysis showed that there were regions where both repeats occurred, as shown in Supplemental Figure S3. This verifies that there are two types of ordering although they are only local and the system is better considered as a surface-solution of different compositions, similar to a bulk solid solution.

**X-ray absorption spectroscopy.** To clarify the structural transition, we performed XAS measurements using the Auger electron yield (AEY) and the secondary electron yield (SEY), which provide fingerprints of the local coordination environments for Ti atoms in the near-surface region. In AEY the x-ray beam excites a photoelectron for a specific state; and the Auger electron emitted when the state decays are detected. As electrons with a kinetic energy of several hundred eV strongly interact with solids, their inelastic mean free path (IMFP) is short.[41] For the specific case of Ti $L_{M,M}$ transitions (450 eV) in grazing exit we estimate the IMFP as 1 nm. This makes XAS-AEY a very surface sensitive technique, where monolayers are easily detectable.[42] In contrast, in SEY the photoelectrons emitted are detected as secondaries leaving the surface in an energy window of 0-20 eV, i.e., with a somewhat larger IMFP. Thus this mode probes layers deeper in the bulk.[43]

Striking differences were observed between the AEY and SEY on the 4×1 surface (see Figure 2(a)). Two main peaks located at 458 and 460 eV correspond to Ti 2$p$3/2→3$d$ transitions, while an additional shoulder appears at 459 eV in the SEY (see the black curve in Figure 2(a)). The shoulder becomes pronounced and the two main peaks are significantly weakened in the AEY (see the red curve in Figure 2(a)). Similar effects were present in the Ti $L_2$ spectra ($2p1/2 \rightarrow 3d$ transitions).[43]

To understand the Ti $L_{2,3}$ spectra, we performed crystal-field multiplet calculations of a $Ti^{4+}$ ion in octahedral ($O_h$) and tetrahedral ($T_d$) symmetry using the CTM4XAS code.[33, 44] As shown in Figure 2(b) the simulated spectra of Ti in $O_h$ and $T_d$ symmetry agree well with previously published experimental spectra for $BaTiO_3$ and $Ba_2TiO_4$ with octahedrally and tetrahedrally coordinated Ti ions, respectively.[45] Very good agreement is achieved between the simulated combination spectra of $O_h$ and $T_d$ and experimental AEY and SEY spectra (see Figure 2(b)). These results verify that the 4×1 surface is composed of tetrahedrally coordinated $TiO_4$ units.

The Ti $L_{2,3}$ XAS spectra on the 2×5 surface show differences between the AEY and SEY data, and significantly differ from the spectra of the 4×1 surface, indicating that the Ti coordination environment changes when the surface structure changes. We note that the spectrum measured with the surface sensitive AEY mode resembles that of a lepidocrocite-like titania nanosheet[46] implying some structural similarity. Furthermore, both valence band and core-level photoemission spectra indicate that the valence of Ti is 4+ and the 2×5 surface is insulating (see Supplemental Figure S4), in line with a surface prepared in an oxygen atmosphere.

**Structural model.** Based upon a structure similar to lepidocrocite, we now turn to an identification of the bright feature in the rows, as well the structure below the outermost layer that is invisible to STM. With full occupancy of all the Ti sites in the second layer, the surface would be a metallic conductor, similar to previous reports of lepidocrocite-like titania as cited earlier. Reducing the second-layer occupancy of titanium leads to a valence-neutral and insulating surface. From Sr deposition experiments, there was a circumstantial link between the bright features and the presence of extra Sr atoms. From a detailed DFT analysis of different occupancies, and coupling this with refinements against the diffraction data, we isolated two homologous series of 2×n structures as shown in Figure 3, together with the 3×1 and 4×1 for reference. The first, 2×na, contains two Sr atoms in the trench every unit cell along the "2" direction with a separation of (0,0,1), while the second 2×nb contains one every other with a (0,0,2) spacing and a bridging $TiO_x$ unit. Both are strictly 2×n structures, although the 2×na family will not appear so in STM images. Careful analysis of the diffraction data (see

Supplemental Figure S5) indicated that the "2×" component was disordered. Due to this disorder the Shelx code[47] was used to refine the diffraction data against a combination of either 2×4a,b or 2×5a,b, i.e. statistical fractional occupancy, yielding a crystallographic R1 of 15% with ~1/4 coverage of 2×4a, and an R1 of 22% with 56 measurements for the 2×5 with also ~1/4 coverage of 2×5a, see Supplemental Tables S1 and S2. While these are high for a bulk structure, they are good numbers for a surface indicating that one can have confidence in the structures.[40] For completeness, the DFT calculations indicate the possibility of additional higher-$TiO_2$ coverage structures without any Sr in the trenches, but since we have no diffraction data to confirm this, we will not pursue these further here.

The DFT convex-hull is shown in Figure 4 with error-bars of 0.1 eV/1×1 cell. This is a conventional representation of the energetics versus composition, here the enthalpy versus surface excess of $TiO_2$. The lower envelope describes the single or two-phase coexistence regimes as a function of surface composition. The 2×na family occur at lower surface excess of $TiO_2$, consistent with coexistence with n×1 reconstructions for the high-temperature samples.[13] At the high $TiO_2$ coverage (right) of the convex hull both 2×n families are converging to a straight line to a nominal 2×∞ structure (see Supplemental CIF file). For a straight line one expects extensive co-existence of structures due to the entropy of mixing similar to that found for the (111) surface,[12] which is consistent with the experimental results.

Turning to the structures, all the atomic coordinates are available in the Supplemental Material as CIF files. The 2×n surface structures are good insulators with a network of $O_h$ coordinated $TiO_6$ or $TiO_5[]$ (where [] is either a very long Ti-O distance or a vacant site) obeying bond-valence criteria.[16] The arrangement of Ti in the second layer follows bonding considerations. In the top layer (visible in the STM images) all the available titanium $O_h$ sites are filled except at the origin, but there is only partial occupancy in the second layer. Based upon exploring all the different possibilities, the occupancies follow some simple rules: (i) all the oxygen sites are filled; (ii) the total number of occupied sites is that to yield valence neutrality of the overall structure; (iii) vacant sites do not want to be adjacent along either the long axis of the structure, or the short axis (iv) vacant sites want to be as far apart as possible.

Somewhat unusual are the octahedral $SrO_5[]$ units, which have coordination similar to that in bulk SrO but with longer bond lengths. Using as a reference the bond-valence sum (BVS) for bulk SrO of 1.73, the Sr atoms at the surface are slightly under-bonded for the structures with only one Sr per 2×n cell with BVS of 1.56 (2×4a) and 1.62, 1.69 (2×5a); and severely under-bonded with double Sr occupancy with BVS of 1.18 (2×4b), 1.21 (2×5b). However, the oxygens to which they are bonded all have reasonable BVS numbers of 1.6-1.8. We note that the strontium is a hard $Sr^{2+}$ ion and is present mainly to satisfy ionic neutrality rather than having true covalent bonding, whereas the oxygens have valence electrons and their reasonable BVS numbers are indicative of appropriate coordination and accommodation of the 2p valence electrons, a necessary condition for a stable surface.

**Discussion**

To understand the difference between the low-temperature and high-temperature experiments, we note that ordering can be relatively slow. While oxygen atoms can diffuse relatively fast, the activation energy barriers for Ti/Sr cation migration is higher. STM is a local probe whereas the electron diffraction measurements are statistical averages over the microns coherence width of the electron beam. There is sufficient disorder in the STM images for the low-temperature samples that diffraction data would show streaking rather than sharp spots. Without diffraction data to delineate what is below the outermost surface for some of the probably metastable surfaces seen by STM in the low-temperature samples, we will not speculate about their detailed structure. With the higher-temperature experiments the convex-hull energies are consistent with the experimental data.

The reason the surface switches from tetrahedral to octahedral coordination is stress- and packing density-driven. The tetrahedral coordination in the n×1 reconstructions is stabilized by large tensile strains[13] which can occur for a single monolayer with a low $TiO_2$ coverage. This is comparable to pseudomorphic growth where the substrate drives the structure of a thin film, here it is driving stabilization of tetrahedral co-ordination. With a higher coverage of $TiO_2$ the stresses and strain energy required to maintain tetrahedral co-ordination become excessive, and the lower energy octahedral coordination will be favored, although this is still strained as evidenced by the underbonding apparent in the BVS numbers. Due to the higher coordination numbers of the oxygen atoms at the surface, the Ti-O bond lengths in the surface plane are longer with approximate spacings along the "n" direction in fractional coordinates of (0,1/7) and (0,1/9) for the 2×4 and 2×5 structures respectively, with the outermost Ti at the positions (0,$m$/14) and (0,$m$/18) respectively with $m$ even, and the second layer partially occupied sites at odd values, leading to a net compressive strain. Stress relief occurs via the formation of a discontinuous essentially 1D reconstruction with a long repeat along the n direction, with additional Sr atoms to satisfy valence neutrality as well as to coordinate with the oxygen atoms at the edges of the rows. These bridging units play the same role of stress relief of the overlayer with respect to the underlying bulk $SrTiO_3$ as misfit dislocations would in classic epitaxial strained layer growth. At higher coverages (> 4 monolayers) the STM images suggest a transition to a metallic structure closer to lepidocrocite titania nanosheets.[24-26] We suspect there are misfit dislocations at the buried interface although the exact Ti atom distribution below the outermost layer accessible to STM imaging is unclear so we will not speculate further here.

These results indicate that the evolution of an oxide thin film from a reconstruction towards a multilayer epitaxial thin film can involve a sequence of structures with competing energetics. This phenomenon probably occurs in many other thin oxide films such as those of interest for oxide electronics. As mentioned in the introduction there is already evidence that controlling these surface structures can be more important than how layers of an oxide structure are deposited by MBE.[17-19] There may well be many complex structural issues which will merit attention as interest develops in producing oxide multilayers with more complex structures with high reproducibility for commercial applications. There are already other indications of this in the literature; for instance, how heteroepitaxial oxide films grown on $SrTiO_3$ (001) can depend upon the initial surface structure [48, 49] as can homoepitaxy.[20]

**Supporting Information**

LEED patterns, surface structure diagram, X-ray photoemission spectra, transmission electron diffraction pattern and two tables supporting the conclusions presented in the main text. This material is available free of charge via the Internet at http://pubs.acs.org.

**Corresponding Author**

*Email: zhiming.wang@psi.ch

**Notes**

The authors declare no competing financial interest.

**Acknowledgments**

LDM acknowledges funding by the DOE on Grant No. DE-FG02-01ER45945; AL and AS funding by the NSF on grant number DMR-1206320; JE support from Northwestern University Institute for Catalysis in Energy Processes (ICEP) on grant number DOE DE-FG02-03-ER15457; BCR, MRC and LDM by the NSF on grant number DMR-0710643. EM acknowledges funding from the Austrian FWF under Project W124 (Solids4Fun). ZM acknowledges support by the Austrian Science Fund (FWF), project F45-07 (FOXSI), and UD acknowledges support by the ERC Advanced Grant 'OxideSurfaces'. UD and MRC acknowledge support from COST Action CM1104. AM acknowledges support by the Natural Sciences and Engineering Council of Canada (NSERC) and the Canada Research Program. RJG acknowledges support by NSERC. The authors acknowledge useful discussions with J. Redinger and F. Mittendorfer, Vienna University of Technology, and help with the XAS measurements by Karina Schulte, MAX IV Laboratory, Lund University.

**Figure and Figure Captions**

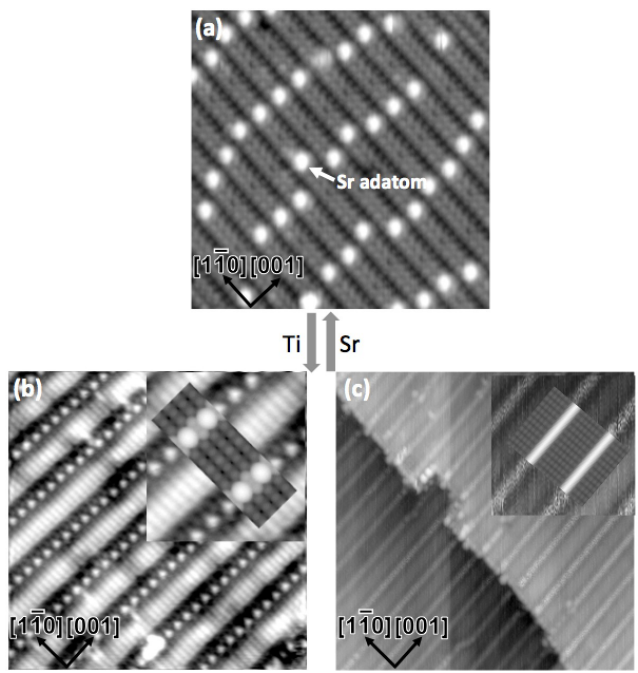

FIG. 1. STM images of the 4×1 (a) and two kinds of 2×5 surfaces (b-c). The arrows in between indicate that these structures can be switched reversibly by depositing Ti/Sr followed by annealing. The Sr single adatom on the 4×1 surface is labeled by an arrow. The insets in (b-c) show magnified views, superimposed with simulated STM images of 2×5a in (b) and 2×5b in (c).

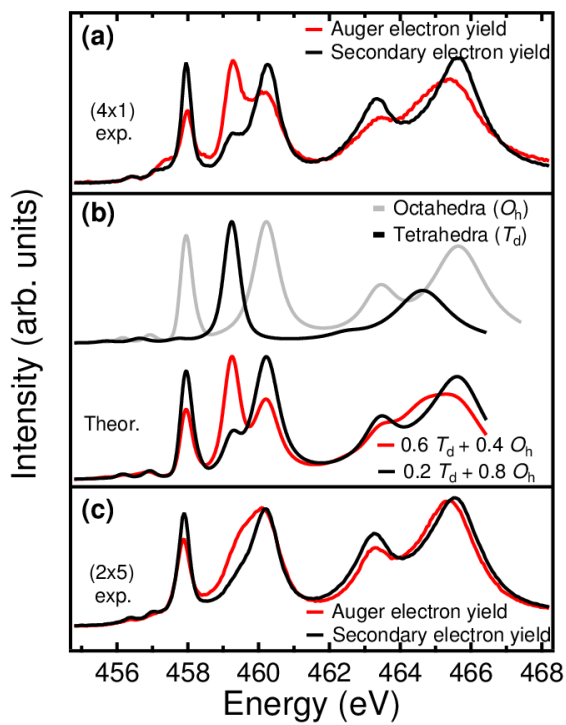

FIG. 2. Ti $L_{2,3}$ XAS on the 4×1 and 2×5 surface. (a) Experimental Ti $L_{2,3}$ XAS measured with AEY (red curve) and SEY (black curve) mode, respectively. (b) Simulated Ti $L_{2,3}$ XAS of Ti ions in octahedra ($O_h$) (grey) and tetrahedra ($T_d$) (black) coordination. The low panel shows the spectra with combination of $O_h$ and $T_d$. (c) Experimental Ti $L_{2,3}$ XAS measured with AEY (red) and SEY (black) electron yield on the 2×5 surface.

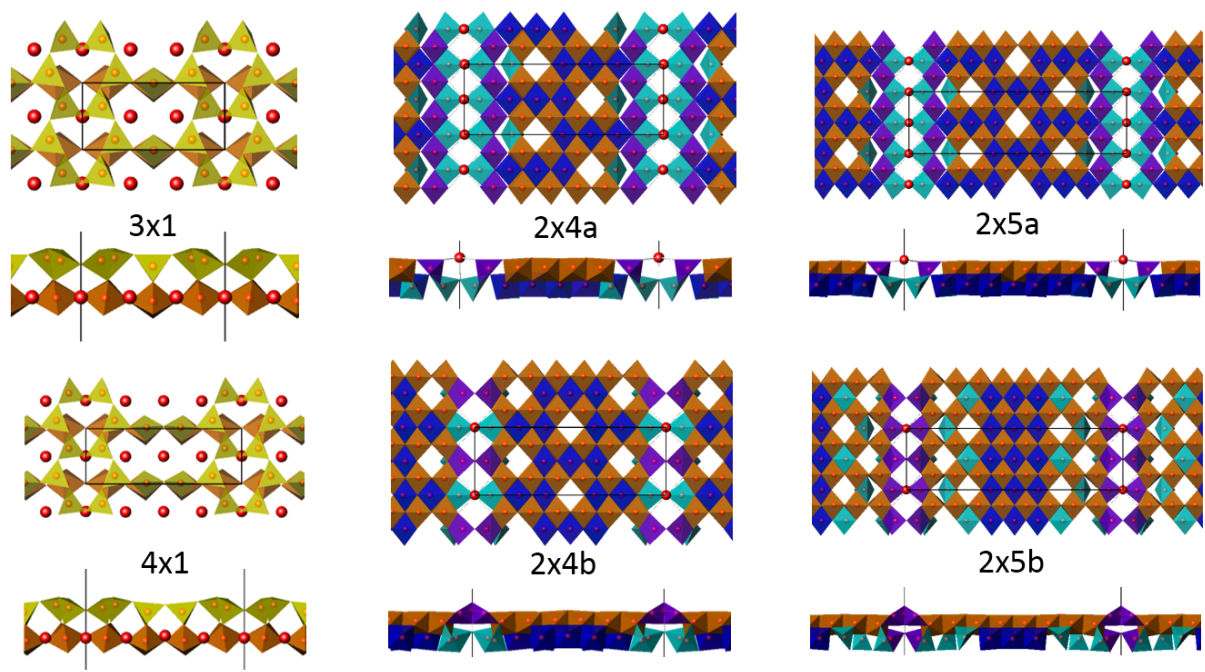

FIG. 3. Polyhedral representations of the structures (color online), top normal to the surface and below from the side. The 3×1 and 4×1 octahedra are brown, tetrahedra golden; in all the others octahedra in the outermost layer are brown, dark blue in the 2$^{nd}$ layer while TiO$_5$[] are purple in the top layer and light green in the second layer.

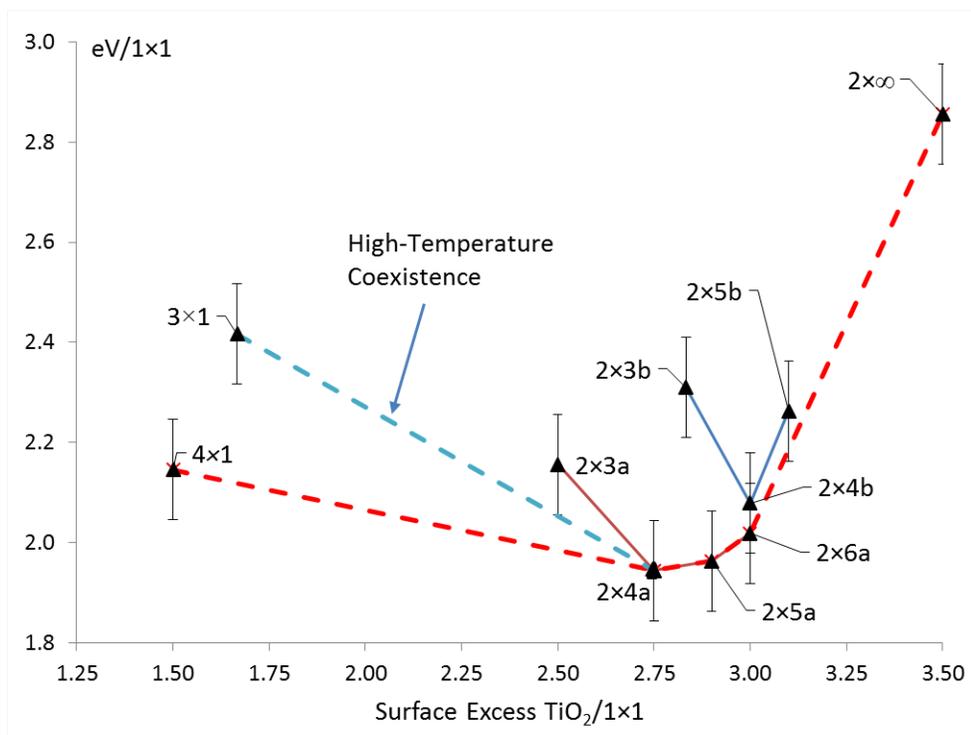

FIG. 4. Convex-hull construction (red) of the surface enthalpies in eV/1×1 (y-axis) for different TiO$_2$ excess per 1×1 unit cell (x-axis), revTP SSh functional. Error bars are 0.1 eV/1×1 cell. Lines are shown for the 2×na and 2×nb families, as well as the 2-phase coexistence (blue dashes) between the 3×1 and and 2×4a structures observed for the samples prepared at higher temperatures.[13]